\documentclass[aps,preprint,showpacs,superscriptaddress,groupedaddress]{revtex4}  
\usepackage{graphicx}  
\usepackage{dcolumn}   
\usepackage{bm}        
\usepackage{amssymb}   
\usepackage{amsmath}
\usepackage{epstopdf}
\usepackage[T1]{fontenc}  
\usepackage{textcomp}     
\usepackage{graphics}
\usepackage{epsfig}
\usepackage{psfrag}
\usepackage{color}
\usepackage[utf8]{inputenc}
\usepackage{scrextend}

\DeclareMathAlphabet{\pazocal}{OMS}{zplm}{m}{n}
\usepackage{float}
\usepackage[section]{placeins}
\newcommand*{\rom}[1]

\begin{document}
\title{$\mu$-hybrid inflation with low reheat temperature and\\ observable gravity waves}
\date{\today}
\author{Mansoor Ur Rehman}
\email[E-mail: ]{mansoor@qau.edu.pk}
\affiliation{Department of Physics, Quaid-i-Azam University, Islamabad 45320, Pakistan}
\author{Qaisar Shafi}
\email[E-mail: ]{shafi@bartol.udel.edu}
\affiliation{Bartol Research Institute, Department of Physics and Astronomy, University of Delaware, Newark, Delaware 19716, USA}
\author{Fariha K. Vardag}
\email[E-mail: ]{f.vardag@gmail.com}
\affiliation{Department of Physics, Quaid-i-Azam University, Islamabad 45320, Pakistan}

\begin{abstract}
In $\mu$-hybrid inflation a nonzero inflaton vacuum expectation value induced by supersymmetry breaking is proportional to the gravitino mass $m_{3/2}$, which can be exploited to resolve the minimal supersymmetric standard model $\mu$ problem. We show how this scenario can be successfully implemented with $m_{3/2} \sim 1-100$~TeV and reheat temperature as low as $10^6$~GeV by employing a minimal renormalizable superpotential coupled with a well defined non-minimal K\"ahler potential. The tensor-to-scalar ratio $r$, a canonical measure of primordial
gravity waves in most cases is less than or of the order of $10^{-6}-10^{-3}$.

\end{abstract}
\pacs{98.80.Cq}
\maketitle
\section{\label{Intro}Introduction}
In its simplest formulation minimal supersymmetric hybrid inflation can be associated with a symmetry breaking $G \rightarrow H$, where $G$ and $H$ are usually assumed to be local gauge symmetries \cite{Dvali:1994ms}. Successful inflation employs a unique renormalizable superpotential $W$ and a canonical K\"ahler potential $K_c$ \cite{Dvali:1994ms, Copeland:1994vg}, and the temperature fluctuation $\delta T/T$ in this case is roughly proportional to $(M/m_P)^2 $\cite{Dvali:1994ms}, where $M$ and $m_P$ denote the symmetry breaking scale of $G$ and reduced Planck scale ($2.4 \times 10^{18}$~GeV), respectively. The model therefore predicts that the symmetry breaking scale $M$ is on the order of $10^{15}-10^{16}$~GeV, which means that grand unified theories (GUTs) are naturally incorporated in supersymmetric hybrid inflation \cite{Senoguz:2003zw}. A scalar spectral index in agreement with the measurement $n_s =0.9655 \pm 0.0062$ by Planck \cite{Ade:2015lrj} is realized by including both the radiative corrections as well as the relevant soft supersymmetry breaking terms in the inflationary potential \cite{Rehman:2009nq}. Without the soft supersymmetry breaking terms, $n_s$ lies close to 0.98, and an alternative way to bring this into agreement with the Planck measurement is to employ a nonminimal K\"ahler potential \cite{BasteroGil:2006cm, urRehman:2006hu}.

An attractive extension of minimal supersymmetric hybrid inflation is the so-called ``$\mu$-hybrid inflation'' \cite{Dvali:1997uq, Okada:2015vka}, which is shorthand for supersymmetric hybrid inflation in the presence of the trilinear coupling $S H_u H_d$ that yields the desired minimal supersymmetric standard model (MSSM) $\mu$ term. The scalar component of the $G$-singlet superfield $S$ acquires, after supersymmetry breaking, a nonzero vacuum expectation value (VEV) proportional to $m_{3/2}$, where $m_{3/2}$ denotes the gravitino mass \cite{Dvali:1997uq}. This can be used, as shown in \cite{Dvali:1997uq}, to resolve the $\mu$ problem encountered in the MSSM. This model has been explored in greater depth in \cite{Okada:2015vka}, with the conclusion that successful $\mu$-hybrid inflation based on the minimal superpotential $W$ and a canonical K\"ahler potential $K_c$ leads to split supersymmetry \cite{ArkaniHamed:2004yi}. The gravitino mass $m_{3/2}$ and the soft supersymmetry breaking masses are predicted to be larger than $5 \times 10^7\ \text{GeV}$, and the reheat temperature after inflation is estimated to be $T_r\gtrsim10^{12}$~GeV \cite{Okada:2015vka}. For a recent discussion on $\mu$-hybrid inflation in no-scale supergravity see \cite{Wu:2016fzp}, and for a discussion including axions see \cite{Lazarides:2016luu}.

In this paper we study an extension of minimal $\mu$-hybrid inflation in which the canonical K\"ahler potential is replaced by a nonminimal $K$ but the renormalizable superpotential $W$ is retained. This will allow us to implement successful $\mu$-hybrid inflation with $m_{3/2}\sim1-100$~TeV and soft scalar masses in the TeV region, compatible with the resolution of the gauge hierarchy problem. The plan of the paper is as follows. In Sec.~II we review $\mu$-hybrid inflation with the minimal canonical K\"ahler potential.  In Sec.~III we study the consistency of $\mu$-hybrid inflation with the gravitino problem and reheat temperature in the case of a nonminimal K\"ahler potential. We discuss the prospect of observing primordial gravity waves in Sec.~IV. In addition, the impact of cosmic strings \cite{Kibble:1976sj}, if present, on the inflationary predictions is also briefly discussed in Sec.~IV. In Sec.~V we summarize our findings.
\section{$\bm{\mu}$-hybrid inflation with minimal K\"ahler potential}
Minimal supersymmetric $\mu$-hybrid inflation employs a canonical K\"ahler potential and a unique renormalizable superpotential $W$ which respects a $U(1)\ R$ symmetry \cite{Dvali:1997uq},
\begin{equation}\label{SP}
W=\kappa S(\Phi \overline{\Phi}-M^2)+\lambda S H_uH_d,
\end{equation}
 where $\kappa$ and $\lambda$ are dimensionless real parameters. The scalar component of $S$, a gauge singlet chiral superfield, plays the role of the inflaton. The dimension-full parameter $M$ represents the nonzero VEV of the conjugate chiral superfields $\Phi$ and $\overline\Phi$ that belong to a nontrivial representation of a gauge group $G$. Since our main goal is to discuss the viability of $\mu$-hybrid inflation with low reheat temperature and TeV scale SUSY breaking masses, we take $G$ to be $U(1)_{B-L}$ \cite{Jeannerot:1997is}, an attractive model for $\mu$-hybrid inflation, and presumably also the simplest. The superpotential $W$ and superfield $S$ have unit $R$ charges, while the remaining superfields are assigned zero $R$ charges. This implies that in the supersymmetric limit the VEV of the scalar component of superfield $S$ is zero. The gravity-mediated supersymmetry breaking yields a nonzero VEV proportional to $m_{3/2}$ for the scalar component of $S$.  Then the last term in the superpotential, $\lambda S H_uH_d$, effectively describes the $\mu$ term with $\mu\sim ( \lambda/\kappa) 
  m_{3/2}$, thereby solving the MSSM $\mu$ problem \cite{Dvali:1997uq}. 

The minimal canonical K\"ahler potential is given by 
 \begin{equation}\label{kahler}
 K_c=  |S|^2+|\Phi|^2+|\overline{\Phi}|^2 +|H_u|^2+|H_d|^2.
 \end{equation}
  Taking into account the well-known radiative corrections \cite{Dvali:1994ms,Coleman;1973}, supergravity (SUGRA) corrections \cite{Linde:1997sj, Senoguz:2003zw} and the soft supersymmetry breaking terms \cite{Senoguz:2004vu,Rehman:2009nq}, the inflationary potential (along the D-flat direction with $|\Phi|=|\overline{\Phi}|=0$ and $|H_u|=|H_d|=0$) to a good approximation is given by
\begin{equation} \label{potential}
V(x)\simeq\kappa^2M^4\Bigg(1+ \frac{\kappa^2 }{8 \pi ^2}F(x)+\frac{\lambda ^2}{4 \pi ^2} F(y)+\frac{1}{2}\Big(\frac{M }{ m_P}\Big)^4 x^4+a\frac{ m_{3/2}}{\kappa M}x\Bigg).
 \end{equation}
 The radiative corrections are described by the function
\begin{align}
F(x)&=\frac{1}{4}[(x^4+1)\ln\big(\frac{x^4-1}{x^4}\Big)+2x^2\ln\Big(\frac{x^2+1}{x^2-1}\Big)+2\ln\Big(\frac{\kappa^2M^2x^2}{Q^2}\Big)-3],
\end{align}
 where  $x=|S|/M$, $y=\sqrt{\gamma}\ x$ and following \cite{Okada:2015vka} we define $\gamma=\lambda/\kappa$ and take the coefficient of the soft linear term $a=-1$ \cite{Rehman:2009nq}. Note that the canonically normalized inflaton field is $\sigma=|S|/\sqrt 2$, where we denote both the superfield and its scalar component by $S$. Note that in this paper we will ignore the imaginary component of $S$ which has previously been analyzed in \cite{Senoguz:2004vu,urRehman:2006hu,Buchmuller:2014epa}. The $\mu$-term coupling $\lambda SH_uH_d$ in Eq.~(\ref{SP}) induces an inflaton decay into Higgsinos with a decay width given by \cite{Okada:2015vka}, 
  \begin{equation}
 \Gamma_S(S \rightarrow \tilde H_u\tilde H_d)=\frac{\lambda ^2 }{8 \pi }m_{S},\\ 
\end{equation}
 where $m_S=\sqrt{2}\kappa M$ is the inflaton mass. The reheat temperature $T_r$ is estimated to be \cite{Kolb:1990vq}
   \begin{equation}\label{reheat}
    T_r\approx\sqrt[4]{\frac{90}{\pi^2{g_*}}} \sqrt{{\Gamma_S } {m_P}},\\
    \end{equation}
    where $g_*$ is 228.75 for MSSM.
     \begin{figure}
\includegraphics[scale=0.45]{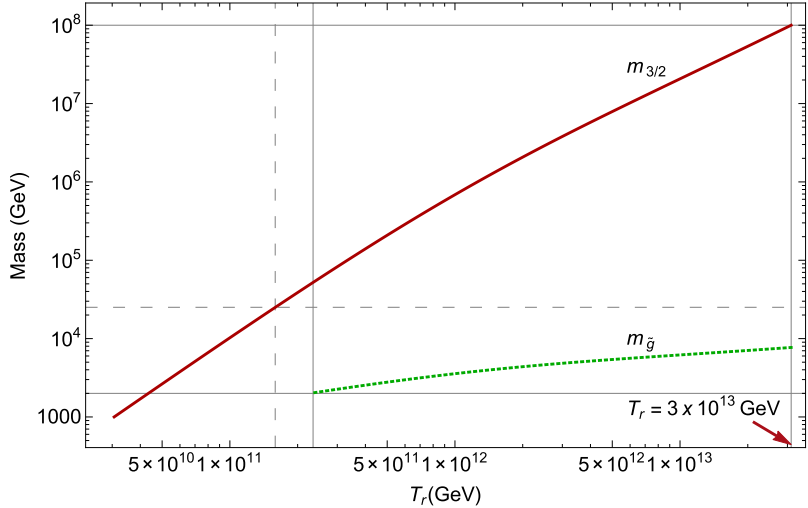}
\caption{\label{fig:M3mgTr}  Variation of gravitino mass $m_{3/2}$ with reheat temperature $T_r$ is shown by a solid-red curve for the minimal K\"{a}hler potential with the conditions, $n_s=0.9655$, $\gamma=2$ and $a=-1$.  The gluino mass $m_{\tilde g}$, obtained via Eq.~(\ref{gluino}), is displayed by a dotted-green curve. The LHC lower bound on gluino mass ($m_{\tilde g}\gtrsim 2$~TeV) is shown by a grey vertical line at $T_r\sim 2\times 10^{11}$~GeV. Regarding the unstable gravitino scenario, we indicate $T_r\sim10^{11}$~GeV corresponding to $m_{3/2}=25$~TeV as shown by a dashed-grey line.} 
\end{figure}

  Following \cite{Okada:2015vka}, we briefly recapitulate $\mu$-hybrid inflation with minimal $K$ and $W$.
  Taking into account the inflationary constraints, we obtain a lower bound on the reheat temperature  $T_r\gtrsim 3\times 10^{10}$ GeV for a gravitino mass $m_{3/2} \gtrsim 1$~TeV (as shown by the solid-red curve in Fig.~\ref{fig:M3mgTr}). Assuming that the gravitino is the lightest supersymmetric particle (LSP), its contribution to the relic density \cite{Bolz:2000fu} is given by
\begin{equation}\label{gluinos}
\Omega_{3/2}h^2 =0.23\Big(\frac{T_r}{10^{10}\text{ GeV} }\Big)\Big(\frac{1\text{ TeV} }{m_{3/2}} \Big)\Big(\frac{m_{\tilde g}}{2\text{ TeV} } \Big)^2,
\end{equation}
where $m_{\tilde g}$ is the gluino mass. The relic density of the observed dark matter (DM) (that is $\Omega_{DM}h^2=0.11$) \cite{Bennett:2012zja} is used to obtain the variation of the gluino mass $m_{\tilde g}$ with $T_r$ and $m_{3/2}$,
 \begin{equation}\label{gluino}
\frac{m_{\tilde g}}{2\text{ TeV}} =\Big[0.46\Big(\frac{10^{10}\text{ GeV} }{T_r}\Big)\Big(\frac{m_{3/2}}{1\text{ TeV} } \Big) \Big]^\frac{1}{2}.
\end{equation}
Using the LHC bound on the gluino mass, the plot of $m_{\tilde g}$ was cut off at 2~TeV. It is clearly seen that the gravitino is heavier than the gluino for all values above the LHC cutoff (Fig.~\ref{fig:M3mgTr}), and hence the former cannot be the LSP.
 
 In light of the unstable gravitino problem \cite{Ellis:1984eq},  we review the two scenarios discussed in detail in \cite{Okada:2015vka}.  For an  unstable gravitino, the lifetime is (see Fig.~1 of \cite{Kawasaki:2008qe})
 \begin{equation}\label{tau}
 \tau_{3/2}\simeq 1.6\times10^4\Big( \frac{1\ \text{TeV}}{m_{3/2}}\Big)^3.\\
 \end{equation}
A long-lived unstable gravitino with lifetime $\tau_{3/2}>1$ sec and mass $m_{3/2}<25$~TeV will decay after big bang nucleosynthesis (BBN). Therefore, successful BBN limits the reheat temperature to $T_r\lesssim3\times(10^5-10^6)$~GeV and $T_r\lesssim2\times10^9$~GeV for gravitino masses of 1~TeV and $10$~TeV, respectively \cite{Kawasaki:2008qe}.  As shown in Fig.~\ref{fig:M3mgTr}, the reheat temperature from inflationary constraints is $3\times10^{10}$~GeV and $10^{11}$~GeV for gravitino masses of 1~TeV and $10$~TeV, respectively. Hence, the possibility of long-lived unstable gravitino is inconsistent with the BBN bounds.

For a short-lived unstable gravitino with $m_{3/2}>25$~TeV, the BBN bound on reheat temperature is unimportant. As the gravitino decays into the LSP neutralino $\tilde \chi_1^0$, we find \cite{Kawasaki:2008qe}
 \begin{equation}\label{relic}
 \Omega_{\tilde \chi_1^0}h^2\simeq 2.8\times10^{11}\times Y_{3/2} \Big(\frac{m_{\tilde \chi_1^0}}{1\text{ TeV}}\Big),
 \end{equation}
 where $m_{\tilde \chi_1^0}$ is the mass  of the lightest neutralino and $Y_{3/2}$ is gravitino yield given by
 \begin{equation}\label{yield}
 Y_{3/2}\simeq2.3\times 10^{-12}\Big(\frac{T_r}{10^{10}\text{ GeV}}\Big).
 \end{equation}
 This relation fits well with the numerical result for a wide range of reheat temperatures, that is $T_r\sim10^5\text{ GeV}-10^{12}$~GeV \cite{Kawasaki:2008qe}. As the LSP neutralino density produced by gravitino decay should not exceed the observed DM relic density, we finally obtain an upper bound on the neutralino mass,
 \begin{equation}\label{masschi}
  m_{\tilde\chi_1^0} \lesssim 18 \Big(\frac{10^{11} \text{ GeV}}{T_r} \Big).
 \end{equation}
 This is inconsistent with the lower limit set on the neutralino mass $m_{\tilde\chi_1^0}\gtrsim18$~GeV \cite{Hooper:2002nq} for reheat temperature $T_r\gtrsim10^{11}$~GeV with $m_{3/2}>25$~TeV (see Fig. \ref{fig:M3mgTr}).
 
  To bypass this constraint, the LSP neutralino is assumed to be in thermal equilibrium during gravitino decay, in which case the neutralino abundance is independent of the gravitino yield. For a typical value of the freeze out temperature, $T_F\simeq 0.05\  m_{\tilde\chi^0_1}$, the gravitino lifetime is estimated to be
 \begin{equation}\label{tau2}
 \tau_{3/2}\lesssim 10^{-11}\text{ sec}\Big( \frac{1\ \text{TeV}}{m_{\tilde\chi^0_1}}\Big)^2.\\
 \end{equation}
Comparing Eqs.(\ref{tau}) and (\ref{tau2}), we obtain a bound on $m_{3/2}$,
 \begin{equation}
 m_{3/2} \gtrsim 10^8\text{ GeV} \Big(\frac {m_{\tilde\chi_1^0}}{2\text{ TeV}}\Big)^{2/3}.
 \end{equation}
We thus arrive at the main conclusion of \cite{Okada:2015vka}, namely that the successful realization of  $\mu$-hybrid inflation with the minimal K\"{a}hler potential requires split supersymmetry with gravitino mass $m_{3/2} \gtrsim 10^8$~GeV. The corresponding  reheat temperature according to this scenario is $ T_r\gtrsim10^{13}$~GeV  (Fig.~\ref{fig:M3mgTr}). In summary,  the gravitino mass is strongly dependent on reheat temperature due to inflationary constraints. In the next section we discuss how these problems are overcome by employing a nonminimal K\"ahler potential. 
\section{\label{NM}$\bm{\mu}$-hybrid inflation with nonminimal K\"{a}hler potenial}
The K\"ahler potential including the relevant nonminimal terms is given by
\begin{equation}\label{kahler}
K=K_c+\kappa_S  \frac{|S|^4}{4m_P^2}+\kappa_{SS} \frac{|S|^6}{6m_P^4}+...  ~ ~~.
\end{equation}
The corresponding scalar potential takes the following form:
\begin{align} \label{nmpotential}
V(x)\simeq\kappa^2M^4\Bigg(1+ \frac{\kappa^2 }{8 \pi ^2}F(x)+\frac{\lambda ^2}{4 \pi ^2} F(y)+\frac{\gamma_S}{2}\Big(\frac{M }{ m_P}\Big)^4 x^4-\kappa_S \Big(\frac{M }{m_P}\Big)^2x^2+a\frac{ m_{3/2}}{\kappa M}x\Bigg),
 \end{align}
where $\gamma_S = 1 + 2 \kappa _S^2-\frac{7 \kappa _S}{2}-3 \kappa _{SS}$. Using this scalar potential, we can obtain the usual slow roll parameters defined by
 \begin{equation}
 \epsilon = \frac{m_{p}^2}{4M^2}\Big(\frac{V'}{V}\Big)^2,\ \eta=\frac{m_{p}^2}{2M^2}\Big(\frac{V''}{V}\Big),\ \xi^2=\frac{m_{p}^4}{4M^4}\Big(\frac{V'V'''}{V^2}\Big).
  \end{equation}
  It is important to note that all derivatives (denoted by $V'$, $V''$, and $V'''$) in the above expressions are with respect to $x$ and not with respect to the canonically normalized field $\sigma =\sqrt2 |S|$.  The scalar spectral index $n_s$, the tensor-to-scalar ratio $r$ and the running of the scalar spectral index $dn_s/d\ln k$, to leading order in slow roll approximation are given by
  \begin{align}\label{para}
  n_s&\simeq 1-6\epsilon+2\eta,\\
  r&\simeq 16\epsilon,\\
  \frac{dn_s}{d\ln k}&\simeq 16\epsilon\eta-24\epsilon^2-2\xi^2.
  \end{align}
  The amplitude of the  power spectrum of scalar curvature perturbation $A_s$ is given by
  \begin{equation}\label{amp}
 A_{s}(k_0)=\frac{1}{6\pi^2}\Big(\frac{M}{ m_P}\Big)^2\Big(\frac{V^3/{V'}^{2}}{m_P^4}\Big)_{x=x_0},
 \end{equation}
 where $A_s(k_0)=2.196\times10^{-9}$ and $x_0$ is the value of the inflaton field at the pivot scale $k_0=0.05\text{ Mpc}^{-1}$ \cite{Ade:2015lrj}.
The number of $e$-folds between the horizon exit (at pivot scale) and the end of inflation is given by
   \begin{equation}
   N_0=2\Big(\frac{M}{m_P}\Big)^2\int_{1}^{x_0}\Big(\frac{V}{V'}\Big)dx.
   \end{equation}
 Assuming a standard thermal history, $N_0$ is related to $T_r$ as
   \begin{equation}\label{n0}
   N_0=53+\frac{1}{3}\ln\Big(\frac{T_r}{10^9\text{ GeV}}\Big)+\frac{2}{3}\ln\Big(\frac{\sqrt\kappa M}{10^{15}\text{ GeV}}\Big),
   \end{equation}
 where $T_r$ is given by Eq.~(\ref{reheat}). The predictions for the various inflationary parameters are estimated numerically using these relations up to second order in the slow roll parameters. For most of the numerical work, we fix the scalar spectral index at the central value of the range given by Planck, $n_s=0.9655 \pm 0.0062$ \cite{Ade:2015lrj}. It is important to mention here that in the minimal K\"ahler potential case with TeV-scale soft SUSY masses \cite{Rehman:2009nq}, we need to take $a=-1$ in order to obtain the red-tilted spectrum within the $1-\sigma$ Planck bounds. Alternatively, for $a=1$ a similar range for $n_s$ can be obtained by taking an intermediate-scale, negative soft mass-squared term for the inflaton \cite{Rehman:2009yj}. However, with the nonminimal K\"ahler potential we can obtain the central value of $n_s$ with TeV-scale soft masses even for $a=1$ \cite{BasteroGil:2006cm, urRehman:2006hu}. The appearance of a negative mass
 term for the inflaton with nonminimal coupling $\kappa_s$ in the potential [Eq.~(\ref{nmpotential})] actually plays a crucial role in realizing successful inflation. Most importantly, it is shown in \cite{urRehman:2006hu} that a low reheat temperature in the standard hybrid inflation can be obtained with appropriately small values of $\kappa$ (see Fig.~10 of \cite{urRehman:2006hu}). Therefore, with smaller values of $\kappa$ and $\lambda$, we expect similar results in $\mu$-hybrid inflation.

   For suitably small $\kappa$ and $\lambda$ values, the last $50-60$ $e$-folds occur in the flat region of the potential with $|S|$ close to $M$ (that is $x\rightarrow1$). Therefore, we can ignore the radiative corrections, and the potential becomes
   \begin{align} \label{nmpotential}
V(x)\simeq\kappa^2M^4\Bigg(1+\frac{\gamma_S}{2}\Big(\frac{M }{ m_P}\Big)^4 x^4-\kappa_S \Big(\frac{M }{m_P}\Big)^2x^2+a\frac{ m_{3/2}}{\kappa M}x\Bigg).
 \end{align}
 Apart from the linear term this potential resembles the model of spontaneous symmetry breaking inflation (SSBI) \cite{Martin:2013tda}. However, near $x\sim1$, the linear term becomes important and develops a local maximum resulting in ``hilltop inflation'' starting near this maximum \cite{Boubekeur:2005zm}. These hilltop-type solutions are a common feature of supersymmetric hybrid inflation models with nonminimal K\"ahler potential \cite{Pallis:2009pq}.
   Near $x\sim1$ Eq.~(\ref{amp}) becomes 
        \begin{equation}
A_s\approx\frac{\kappa^2}{6\pi^2}\Big(\frac{M}{m_{P}}\Big)^6\Bigg(4\gamma_S\Big(\frac{M }{ m_P}\Big)^4 -2\kappa_S \Big(\frac{M }{m_P}\Big)^2+a\frac{ m_{3/2}}{\kappa M}\Bigg)^{-2}.\\
  \end{equation}
    \begin{figure}
  \includegraphics[scale=0.46]{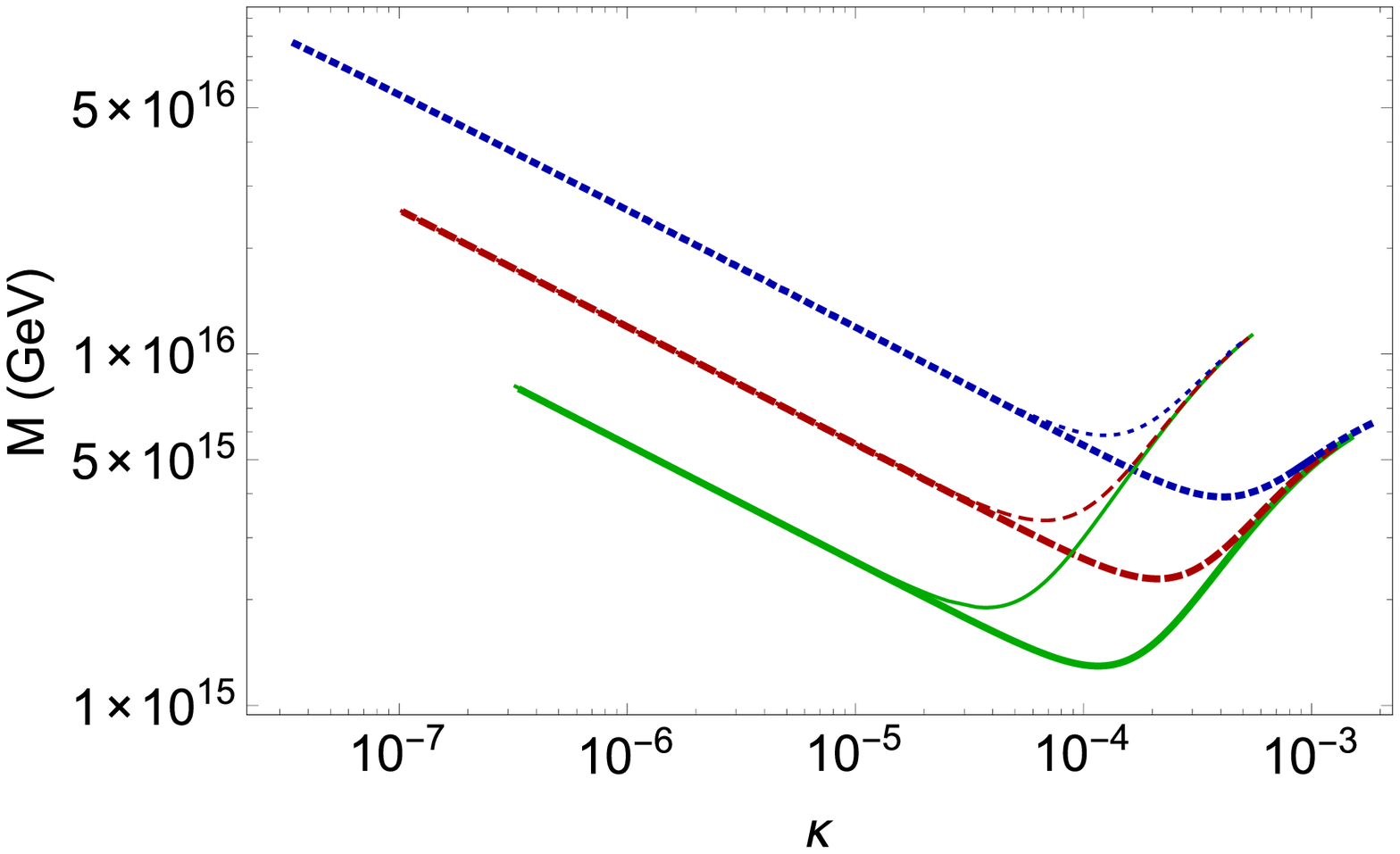}
\includegraphics[scale=0.45]{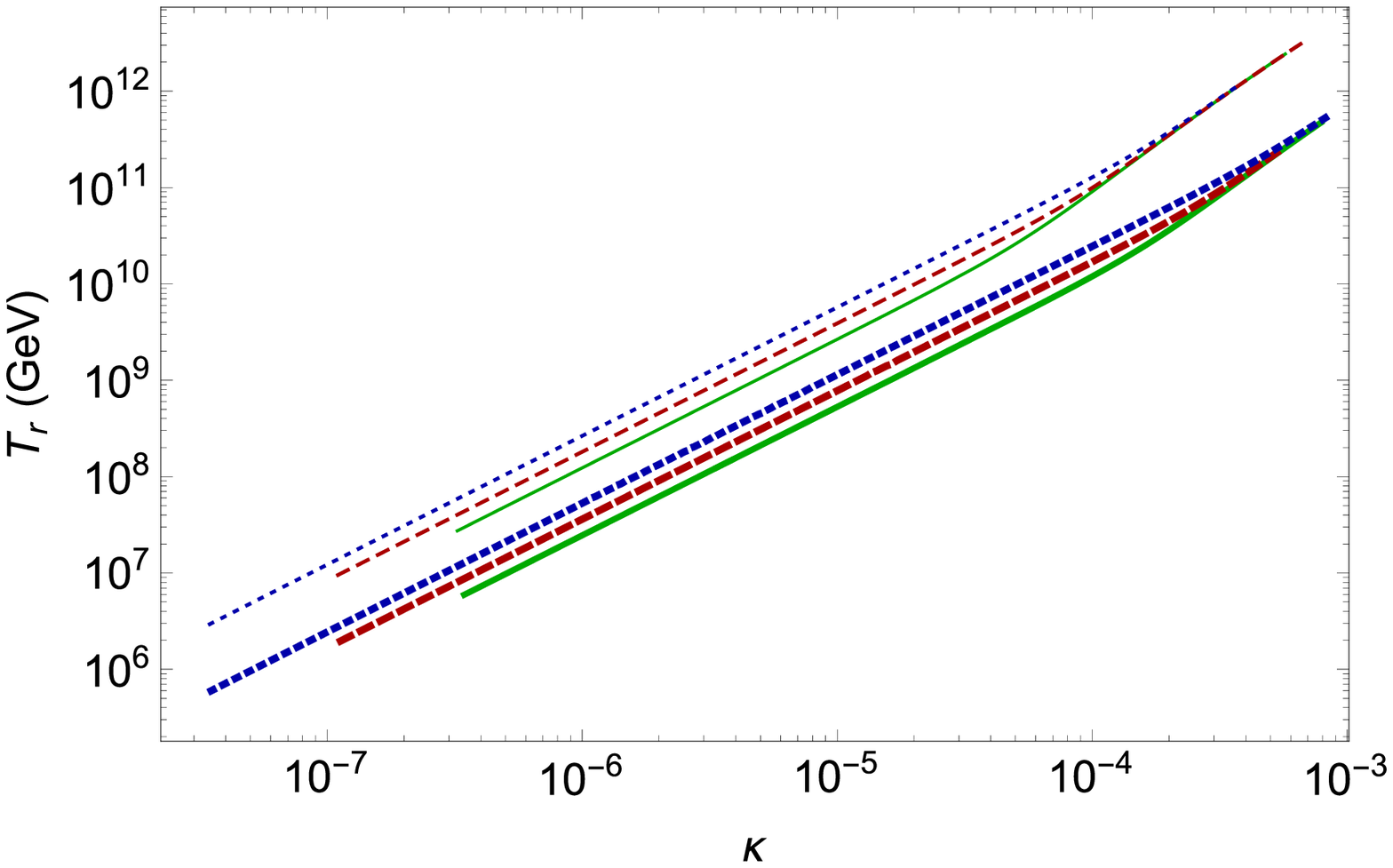}

\caption{\label{fig:Tr2}
 Variation of the reheat temperature $T_r $ and the symmetry breaking scale $M$ with $\kappa$ for gravitino mass of $1$~TeV (solid-green), 10~TeV (dashed-red), and 100~TeV(dotted-blue). We fix the scalar spectral index $n_s=0.9655$, $\kappa_S=0.02$,  $\kappa_{SS}=0$ and represent  $\gamma=2 (10)$ by thick (thin) curves.} 
\end{figure}
Assuming $M\lesssim 10^{16}$~GeV, the quartic SUGRA correction term with coefficient $\gamma_S$ can be ignored and this further simplifies the above expression to
  \begin{equation}\label{As}
A_s\approx\frac{\kappa^2}{6\pi^2}\Big(\frac{M}{m_{P}}\Big)^6\Bigg( -(1-n_s)\Big(\frac{M }{m_P}\Big)^2+a\frac{ m_{3/2}}{\kappa M}\Bigg)^{-2},\\
  \end{equation} where we have used Eq.~(\ref{para}) to eliminate $\kappa_S$ in favour of $n_s$ \cite{urRehman:2006hu}. 
  For  the central value $n_s=0.9655$ and other inflationary constraints from Eqs.~(\ref{amp}) and (\ref{n0}), it is found numerically that there is one percent cancellation among the two terms in Eq.~(\ref{As}). As a result we arrive at a simple relation for $M$ as function of $\kappa$,  
    \begin{equation}\label{mk}
   M(\kappa)\approx \Big(\frac{am_{3/2}}{\kappa(1-n_s)m_P}\Big)^{1/3}m_P\approx5.6\times10^{12}\Big(\frac{m_{3/2}}{\kappa}\Big)^{1/3}\propto\kappa^{-1/3}. 
    \end{equation}
 The reheat temperature from Eq.~(\ref{reheat}) can be written as
  \begin{equation}\label{T}
 T_r (\kappa)\approx1.7\times 10^8\gamma\sqrt{M \kappa^3}\approx3.9\times10^{14} \gamma(m_{3/2})^{1/6}\kappa^{4/3}\propto \kappa^{4/3}.
 \end{equation}
 This is an important relation which justifies our expectation of realizing a low reheat temperature with appropriately small values of $\kappa$ in $\mu$-hybrid inflation. The approximate results in Eqs.~(\ref{mk}) and (\ref{T}) are clearly in agreement with our numerical work as shown in Fig.~\ref{fig:Tr2}. Here we have displayed the variation of $T_r$ and $M$ as a function of $\kappa$, for three different values of the gravitino mass $m_{3/2}=1, 10\text{, and }100\text{ TeV}$.  As mentioned in \cite{Dvali:1997uq}, in order to have the ``good'' point, that is $H_u=H_d=0 \text{ and } \Phi\ \overline{\Phi}=M^2$, as the unique local minimum of the superpotential [given by Eq.~(\ref{SP})], we should have $\gamma>1$. Therefore, to see the effect of $\gamma$  on various inflationary parameters  in general and on reheat temperature in particular, we consider $\gamma=2 \text{ and }10$ (as shown by the thick and thin curves, respectively, in Fig.~\ref{fig:Tr2}). For the $M$ versus $\kappa$ curve of Fig.~\ref{fig:Tr2}, the dependence of $\gamma$ disappears in the low reheat temperature regime where the radiative corrections become unimportant [see Eq.~(\ref{mk})].  For a given value of $\kappa$ and $M$ the reheat temperature varies linearly with $\gamma$  as derived in Eq.~(\ref{T}) and shown in the right panel of Fig.~\ref{fig:Tr2}. Therefore, to estimate the lower bounds on reheat temperature, we implicitly assume $\gamma=2$. As mentioned earlier, the low reheat temperature values occur with $x_0$ close to 1, so we impose a $0.01\%$ fine-tuning bound on the difference $x_0 -1$. Consequently, we obtain the following lower bounds on the reheat temperature $T_r$:
\begin{equation}\label{tuning}
T_r\gtrsim(6\times 10^6,\ 2\times10^6,\ 6\times10^5)\text{ GeV\ \ for\ \ }m_{3/2}=(1, 10, 100)\text{ TeV,}
\end{equation}
respectively (Fig.~\ref{fig:Tr2}).
 \begin{figure}
\includegraphics[scale=0.5]{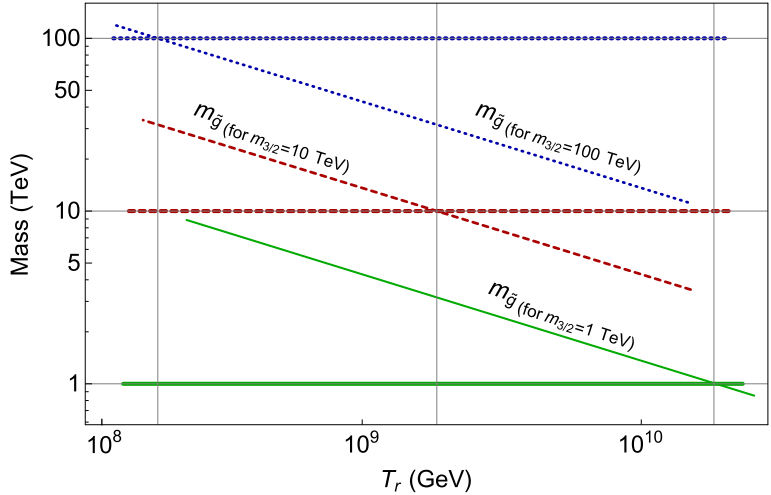}
\caption{\label{fig:mgm32all} Possibility of LSP gravitino with masses $m_{3/2}=(1$, $10$, $100$)~TeV and corresponding gluino masses $m_{\tilde g}$ are shown by solid-green, dashed-red, and  dotted-blue curves, respectively, with $n_s=0.9655$ and $\gamma=2$. The lower bounds on reheat temperature are shown by vertical grey lines.}
\end{figure}

We return to a discussion of the gravitino problem in the light of low reheat temperature. With reference to the previous discussion we consider the following three scenarios:
 \begin{enumerate}
 \item{A stable LSP gravitino;}
 \item{an unstable `long-lived' gravitino with mass less than $25$ TeV;} 
 \item{an unstable `short-lived' gravitino with mass greater than $25$ TeV.}
 \end{enumerate}
 Assuming the gravitino is LSP and using the dark matter relic density ($\Omega_{DM}h^2=0.11$) constraint, the gluino mass is calculated as a function of $T_r$  via Eq.~(\ref{gluinos}), as shown in Fig.~\ref{fig:mgm32all}, for three values of $m_{3/2}$. The upper bounds on the reheat temperature, shown by the vertical grey lines in Fig.~\ref{fig:mgm32all} are
 \begin{equation}
T_r\lesssim 2 \times(10^{10},10^9,10^8)\text{ GeV\   for }m_{3/2}= (1,10,100)\text{ TeV,}
\end{equation} 
respectively.
These upper bounds of $T_r$ are consistent with the lower bound incorporated in our numerical work [i.e. Eq.~(\ref{tuning})], and so the LSP gravitino scenario can be consistently realized in the nonminimal K\"{a}hler case.

 For the second possibility, namely an unstable ``long-lived''  gravitino (with $m_{3/2}\lesssim 25$~TeV), the gravitino problem arises as the gravitino decays right after BBN, which can adversely effect the light nuclei abundances and thereby ruin the success of BBN.  To avoid this problem, as mentioned in Sec. \ref{Intro}, an upper bound on the reheat temperature is obtained \cite{Kawasaki:2008qe}: $T_r \lesssim 3\times(10^5-10^6)$~GeV for $m_{3/2} \sim 1$~TeV, and $T_r \lesssim 2\times10^9$~GeV for $m_{3/2} \sim 10$~TeV. However, the inflationary constraints on the reheat temperature mentioned in Eq.~(\ref{tuning}) are $T_r\gtrsim 6 \times 10^6$ GeV for $m_{3/2}=1$~TeV, and $ T_r \gtrsim 2\times 10^6$~GeV for $m_{3/2}=10$~TeV, respectively. Thus, the second scenario with a long-lived gravitino is marginally ruled out for 1~TeV gravitino mass but stays comfortably within the BBN bounds for a 10~TeV gravitino mass.
 
 An unstable gravitino of mass $m_{3/2}=100$~TeV falls in the third category of a short-lived gravitino. Here the BBN constraints are weakened because a heavy gravitino implies a shorter lifetime. However, now the constraints from the abundance of  the lightest LSP neutralino from the decay of the gravitino will come into play.  These constraints are less severe than those from BBN \cite{Kawasaki:2008qe}. For a 100~TeV gravitino mass, the upper bound on the LSP neutralino as derived in Eq.~(\ref{masschi}) becomes
    \begin{equation}
  m_{\tilde\chi_1^0} \lesssim (18-10^5)\text{ GeV for \ } 10^{11}\text{ GeV}\gtrsim T_r\gtrsim 6\times10^5 \text{ TeV}.
 \end{equation}
 The lower limit on the LSP neutralino mass from experiments is $m_{\tilde\chi_1^0}\gtrsim18$~GeV \cite{Hooper:2002nq}. Therefore, the non-LSP gravitino scenario with $m_{3/2} \sim 100$~TeV comfortably holds in a considerably larger domain: $10^{6} \text{ GeV} \lesssim T_r \lesssim10^{11}$~GeV.
 Hence, we have successfully realized $\mu$-hybrid inflation with $m_{3/2}\sim1-100$~TeV and low reheat temperature.
 \begin{figure}
\includegraphics[scale=0.45]{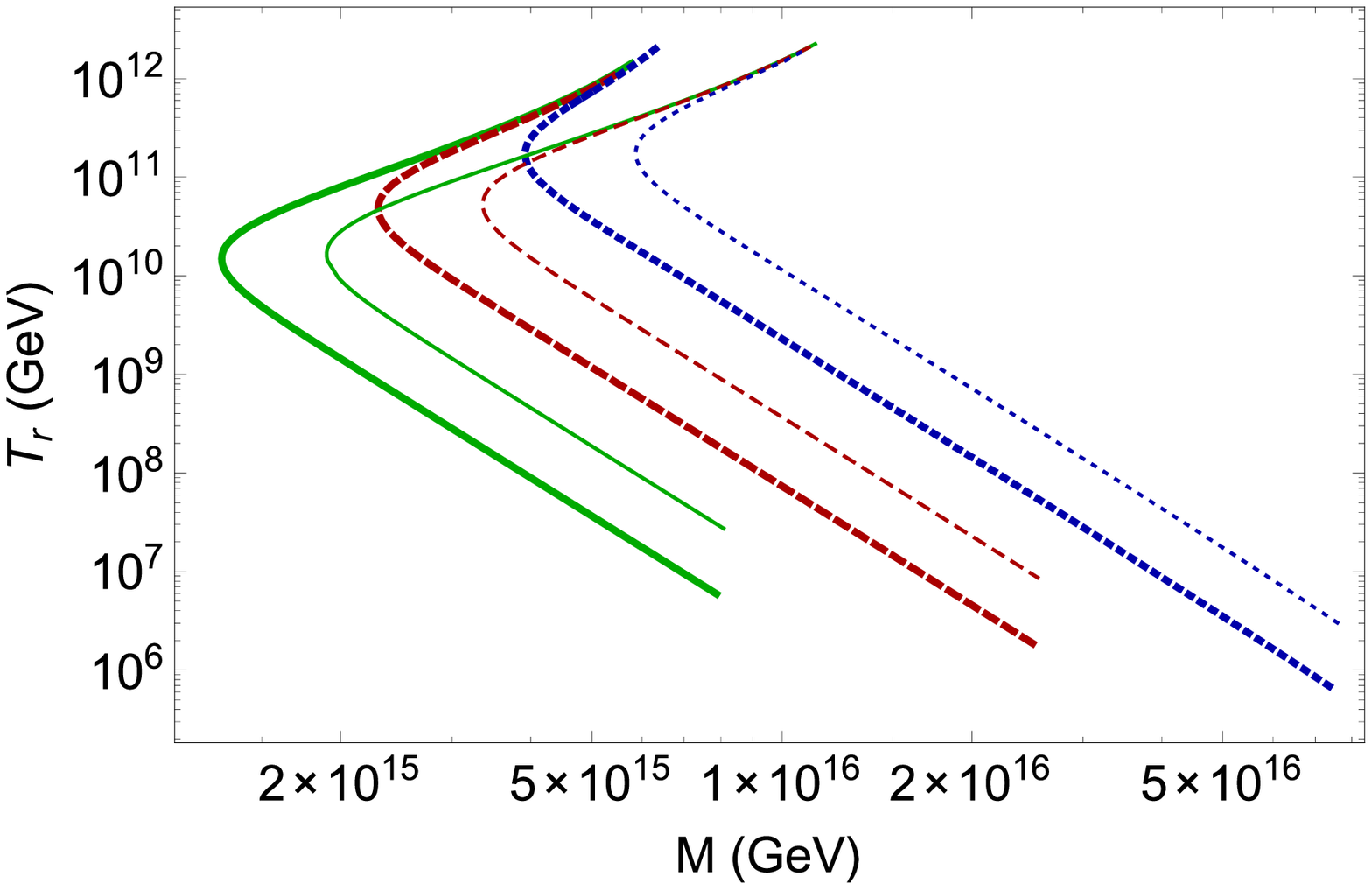}
\includegraphics[scale=0.475]{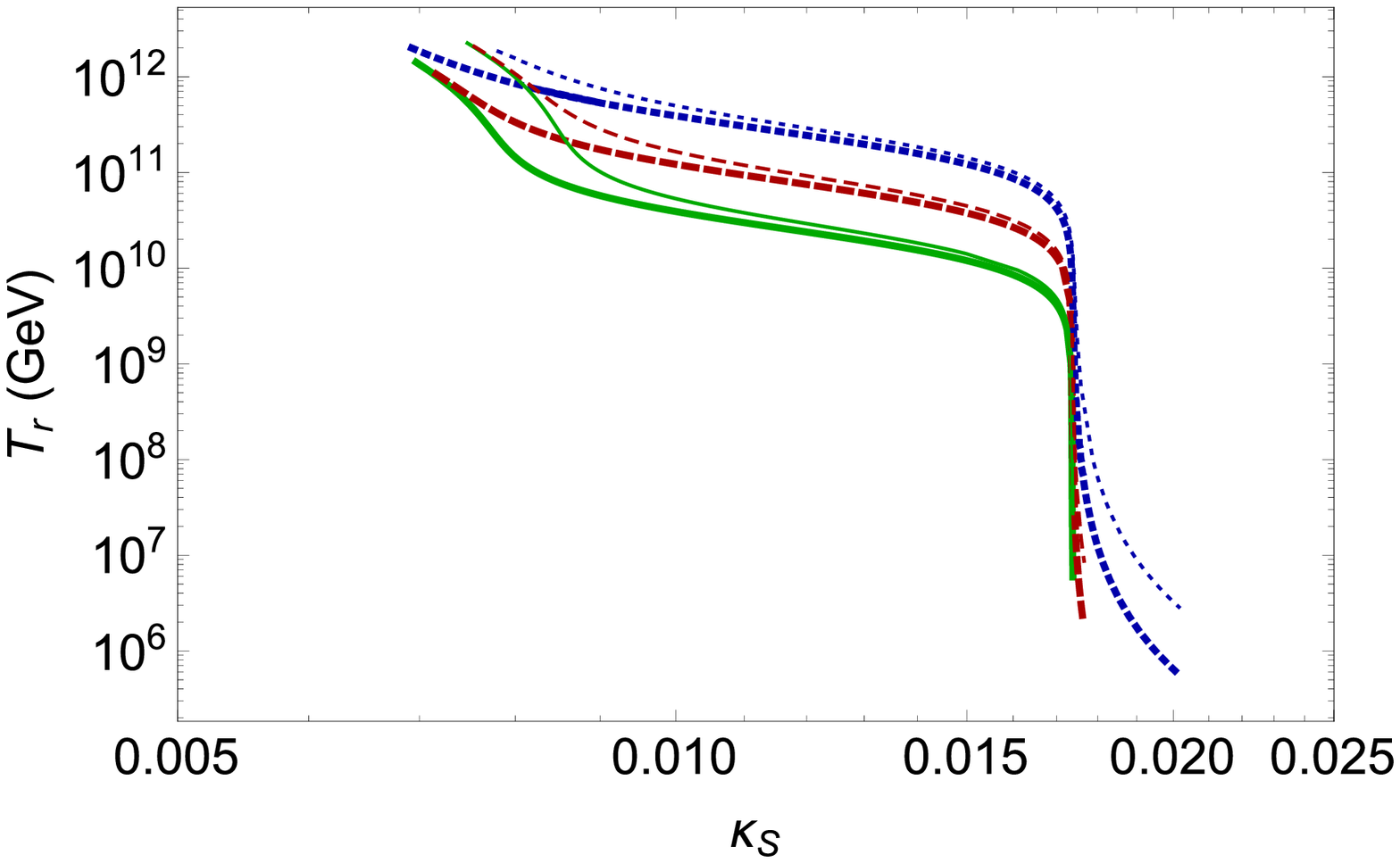}
\includegraphics[scale=0.45]{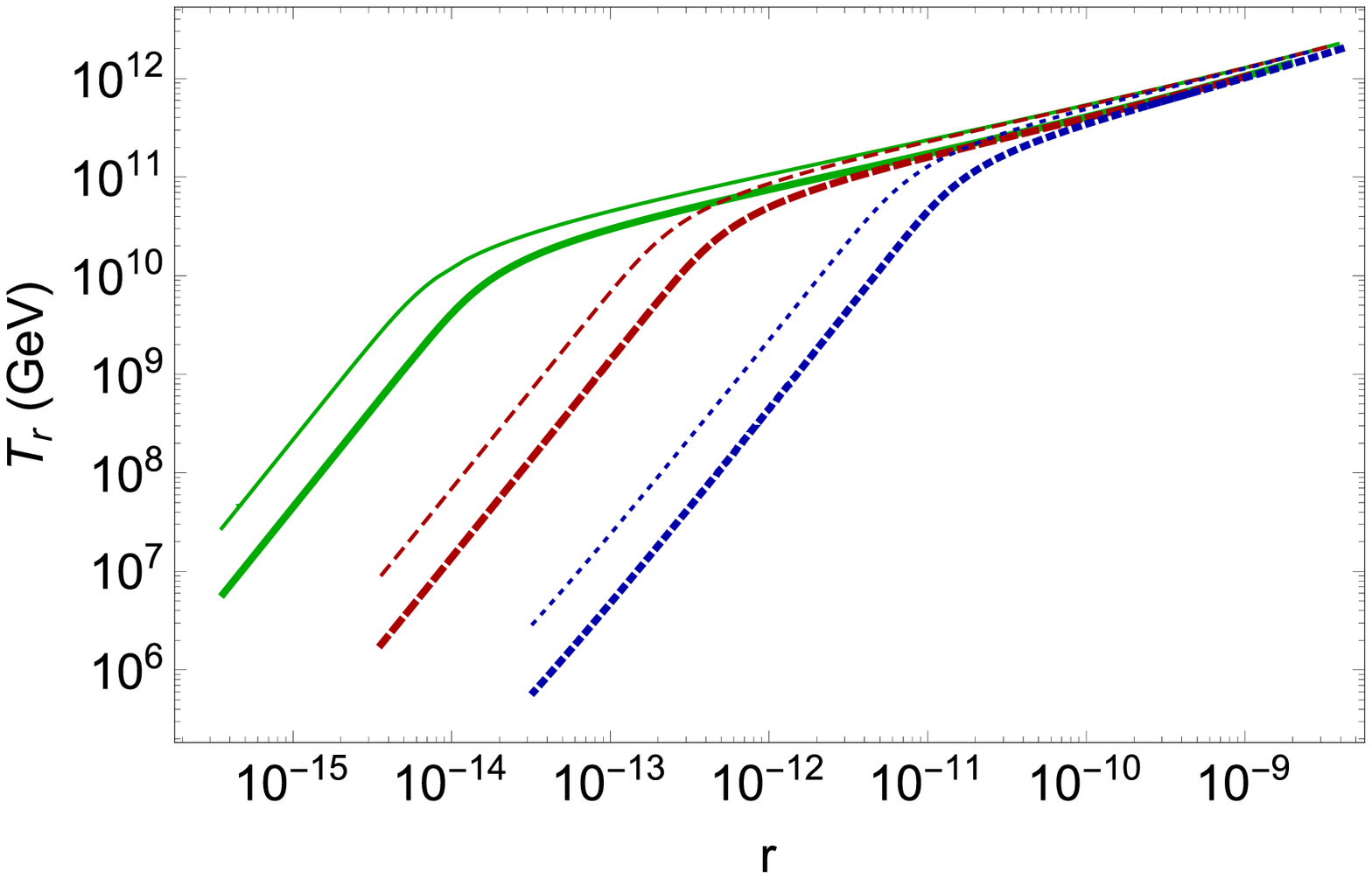}
\includegraphics[scale=0.45]{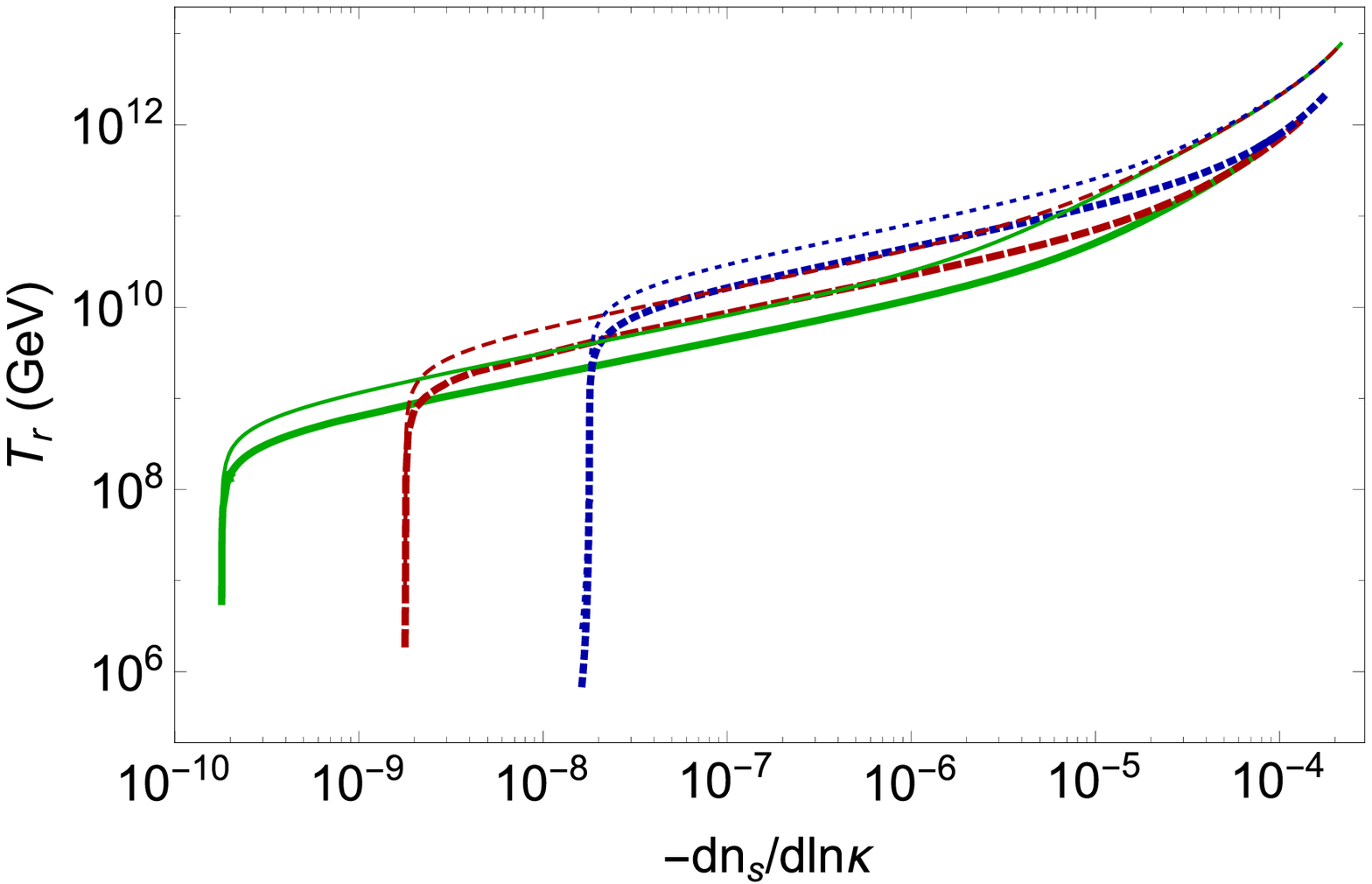}
\caption{\label{fig:Tr}
Variation of the reheat temperature $T_r $ with the symmetry breaking scale $M$, coefficient $\kappa_S$, the tensor-to-scalar ratio \textit{r}, and the running of spectral index $dn_s/d\ln k$, for $m_{3/2}=1$~TeV (solid-green), 10~TeV (dashed-red), and 100~TeV(dotted-blue). We have set the scalar spectral index $n_s=0.9655$, $\kappa_S=0.02$,  $\kappa_{SS}=0$ and represent $\gamma=2 (10)$ by thick (thin) curves.}
\end{figure}

The predictions of other important inflationary parameters are displayed in Fig.~\ref{fig:Tr}. It is interesting to note that with low reheat temperature we can a achieve gauge symmetry breaking scale $M$ of the order of the GUT scale, namely $2\times 10^{16}$~ GeV. In the second panel of Fig.~\ref{fig:Tr}, the value of the parameter  $\kappa_S$ is seen to remain constant in the low reheat temperature range. As discussed earlier, the radiative and the quartic-SUGRA corrections can be neglected in this region. Therefore, the scalar spectral index $n_s$ in the low reheat region  can be approximated as
\begin{equation}
  n_s\simeq 1-2\kappa_S \implies \kappa_S=\frac{1-n_s}{2}.
  \end{equation}
For the central value of scalar spectral index  $n_s= 0.9655$, we obtain $\kappa_S=0.0173$  in good agreement with the numerical estimate shown in Fig.~\ref{fig:Tr}.
The tensor-to-scalar ratio $r$ and the running of the scalar spectral index $dn_s/d\ln k$ in the low reheat limit are given by
   \begin{align}
 r & \lesssim \frac{2}{3\pi^2\sqrt{\gamma/2} A_s(k_0)}\Big(\frac{m_{3/2}}{(1-n_s)m_P}\Big)^{5/4}\Big(\frac{T_r}{m_P}\Big)^{1/2} , \\ 
dn_s/d\ln k &\lesssim -\left(\frac{6}{\pi^2A_s(k_0)} \right)^{1/2} \frac{m_{3/2}}{(1-n_s)m_P}.
  \end{align}
As shown in the lower two panels of Fig.~\ref{fig:Tr}, $r$ and $dn_s/d\ln k$ in this case are very small which is a common feature of small field models. 
\section{Primordial gravity waves and cosmic strings}
\begin{figure}
\includegraphics[scale=0.625]{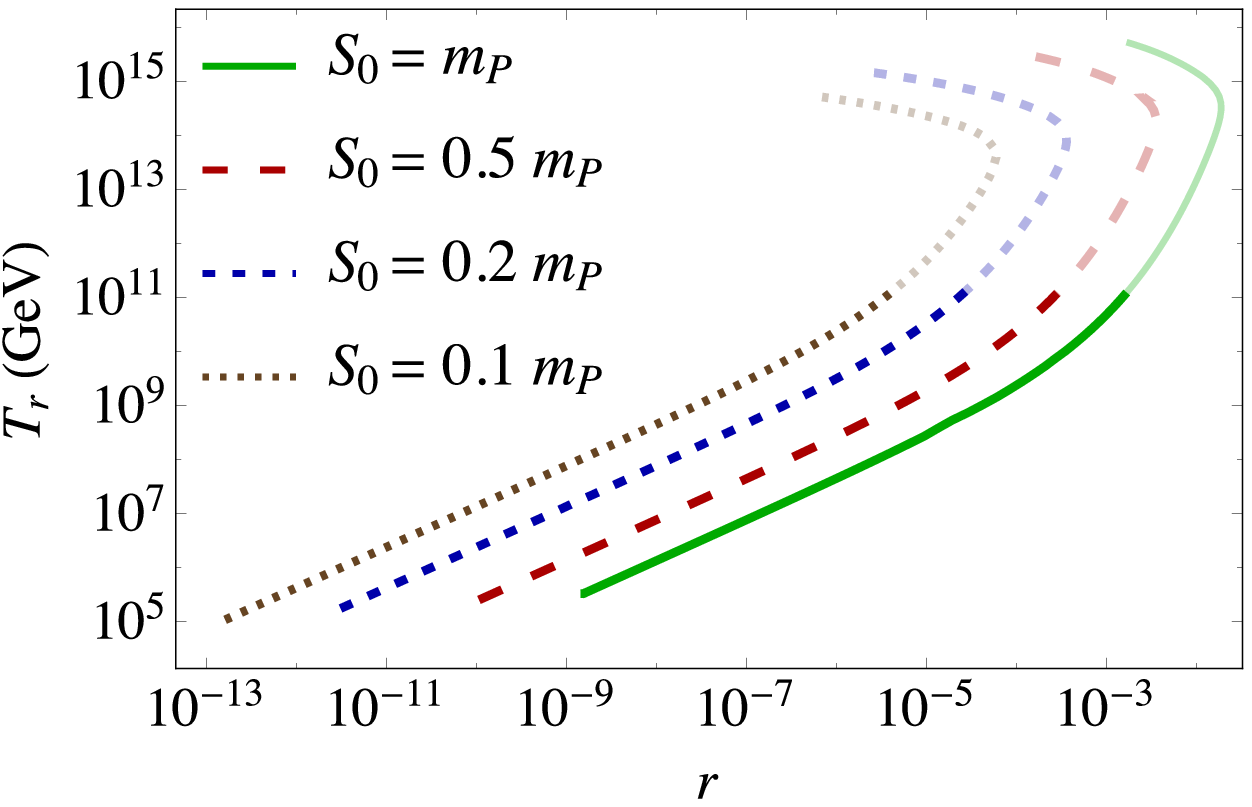}
\includegraphics[scale=0.625]{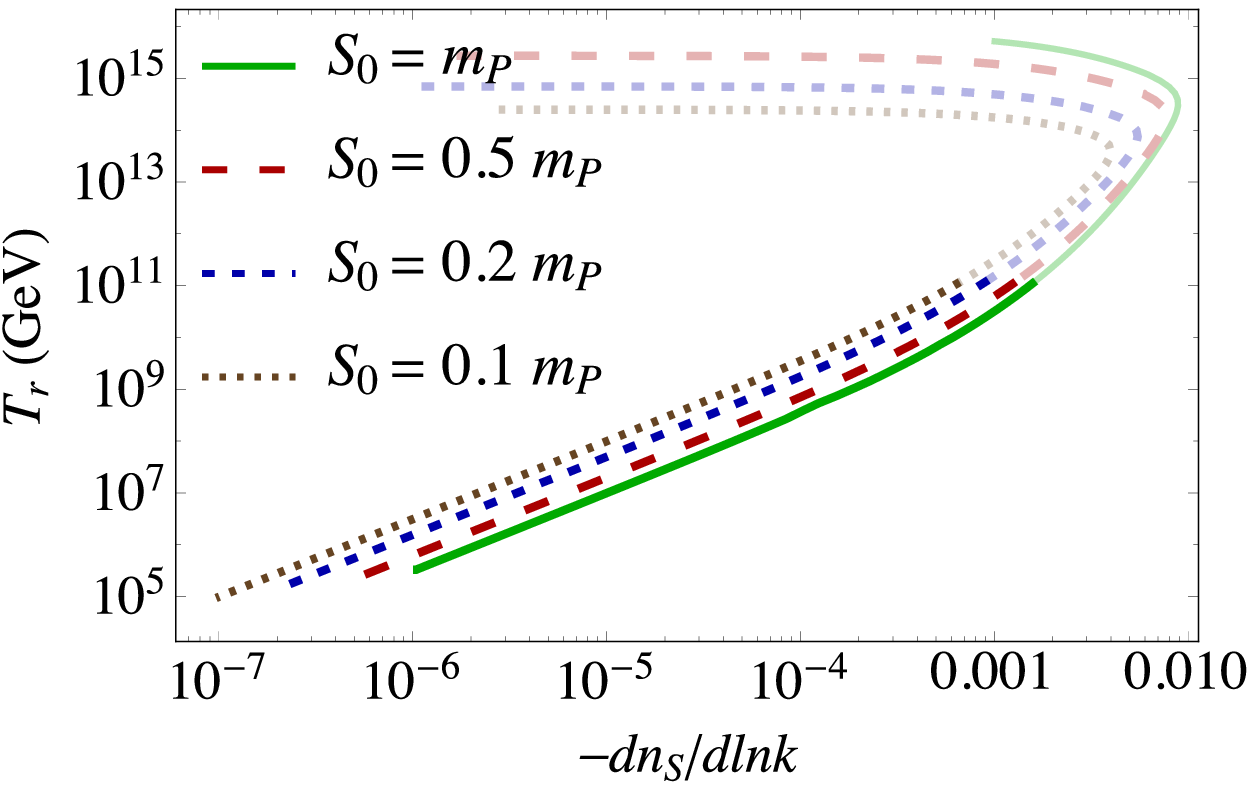}
\includegraphics[scale=0.66]{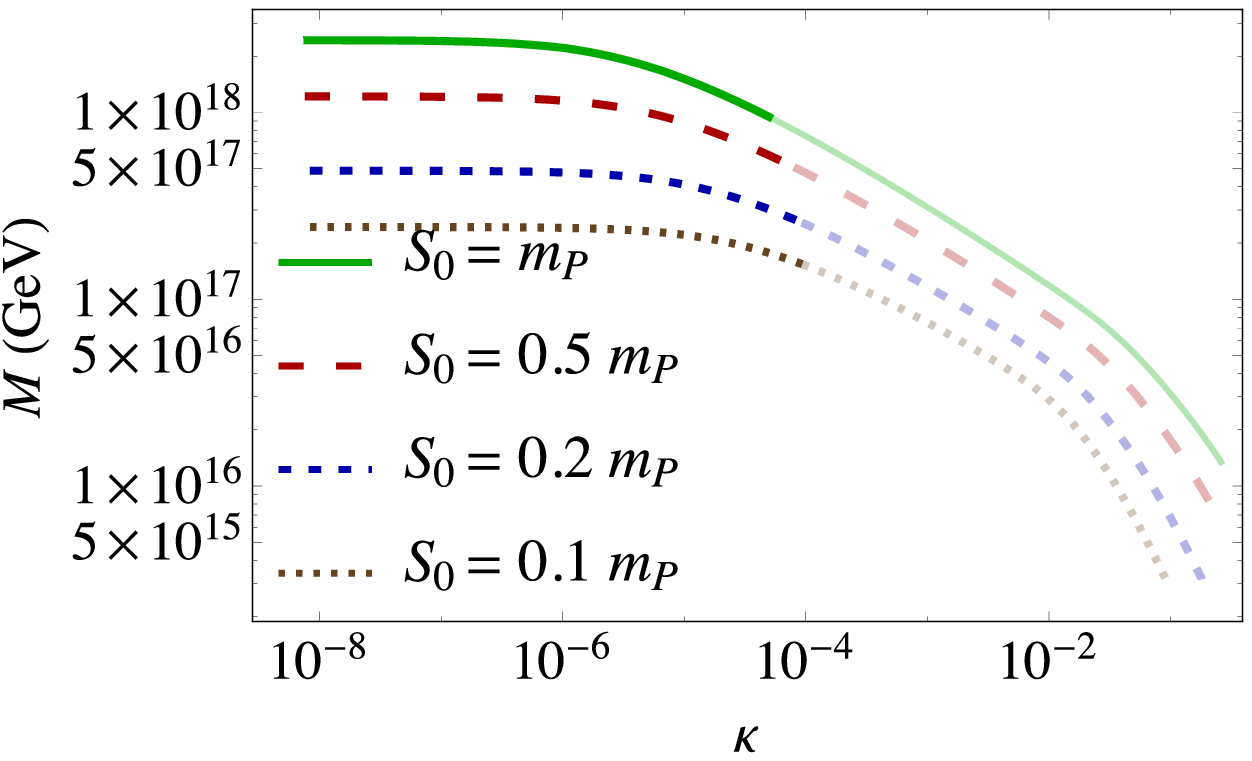}
\includegraphics[scale=0.625]{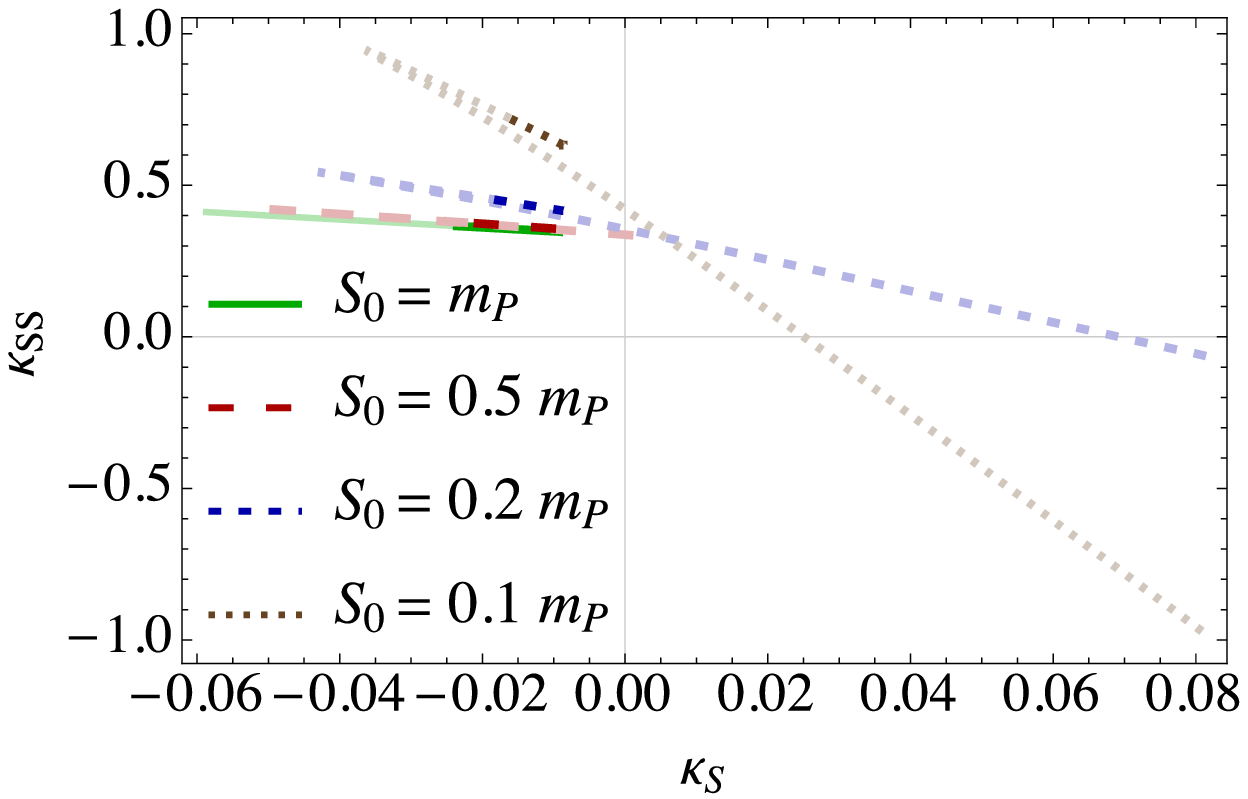}
\caption{\label{fig:larger} The variation of the reheat temperature $T_r$ with $r$ and $dn_s/d\ln k$ (upper panel) and $M$ versus $\kappa$ and $\kappa_{SS}$ versus $\kappa_S$ (lower panel), for $S_0 = m_P$ (green), $m_P/2$ (red), $m_P/5$ (blue), and $m_P/10 $ (brown), valid for gravitino mass range $m_{3/2}\sim1-100$ TeV with $n_s=0.9655$ and $\gamma=2$. The faded regions of the curves represent the reheat temperature $T_r\gtrsim 10^{11} \text{ GeV}$.} 
\end{figure}

So far we have set the nonminimal coupling $\kappa_{SS}$ to be zero in our calculations and only the nonminimal coupling $\kappa_S$ has played a significant role in realizing $\mu$-hybrid inflation. Next, with nonzero $\kappa_{SS}$ we will mainly be interested in the large (observable) $r$ solutions which might be detectable in future experiments: PRISM will be able to measure the tensor-to-scaler ratio $r\lesssim5\times10^{-4}$ \cite{Andre:2013afa}, whereas LiteBIRD will provide a precision of $\delta r<0.001$  \cite{Matsumura:2013aja}. The large $r$ solutions have previously been explored in standard hybrid inflation in \cite{Shafi:2010jr}  with intermediate-scale soft masses, and in \cite{Rehman:2010wm} with TeV-scale soft masses. As explained in these references, large $r$ solutions are possible with positive quadratic and negative quartic terms in the potential. This structure is not possible with $\kappa_{SS}=0$, where, with $\kappa_S=0.0173$, the quadratic term is negative and the quartic term is positive. This structure of the potential resembles the SSBI model of \cite{Martin:2013tda} if we ignore both the radiative corrections and the linear term in $V(x)$, Eq.~(\ref{nmpotential}). It turns out that although the linear term can be ignored for large $r$ solutions, the radiative corrections provide a contribution that is comparable with the quadratic and quartic terms. Therefore, the predictions of our model are, in general, different from those of SSBI model but it lies within the same category of hilltop inflation \cite{Boubekeur:2005zm}. The largest possible values of $r$ consistent with the Lyth bound in our case are obtained with $S_0$ values comparable to $m_P$.  Therefore, we take $S_0=(0.1$, $0.2$, $0.5$, $1)~ m_P$ and show in Fig.~\ref{fig:larger} the variation of reheat temperature $T_r$ with $r$ and $dn_s/d\ln k$ in the upper panel, and $M$ versus $\kappa$ and $\kappa_{SS}$ versus $\kappa_S$ in the lower panel.  To avoid the gravitino problem the bound on reheat temperature $T_r\lesssim10^{11}\text{\ GeV}$ is highlighted by the darkening of curves, whereas the faded region of the curves show large $r$  solutions that lie outside this bound.

The choice $S_0=m_P$ gives the upper bound on $r\lesssim 0.002$ mainly from the requirement of $T_r\lesssim10^{11}\text{\ GeV}$. However, this choice entails the fine-tuning of higher-order terms in the potential. Therefore, with $S_0\sim0.1$~$m_P$, we obtain an upper bound $r\lesssim 4\times10^{-6}$  with reasonably natural suppression of higher-order terms. This is also consistent with the value of $r$  obtained by requiring the boundedness of the potential for large values of the field  as discussed in \cite{Civiletti:2014bca}. 
\begin{figure}
\includegraphics[scale=0.485]{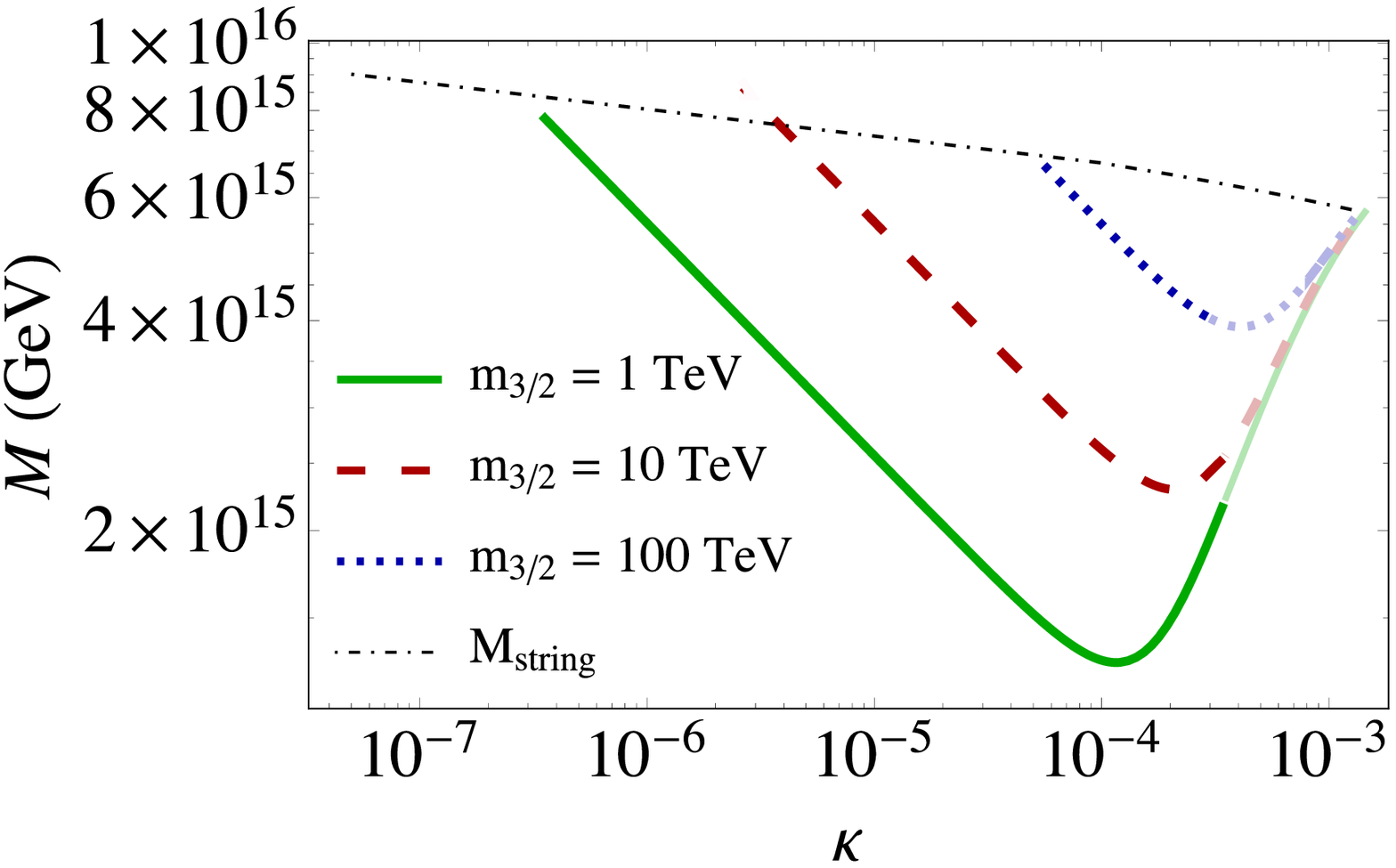}
\includegraphics[scale=0.61]{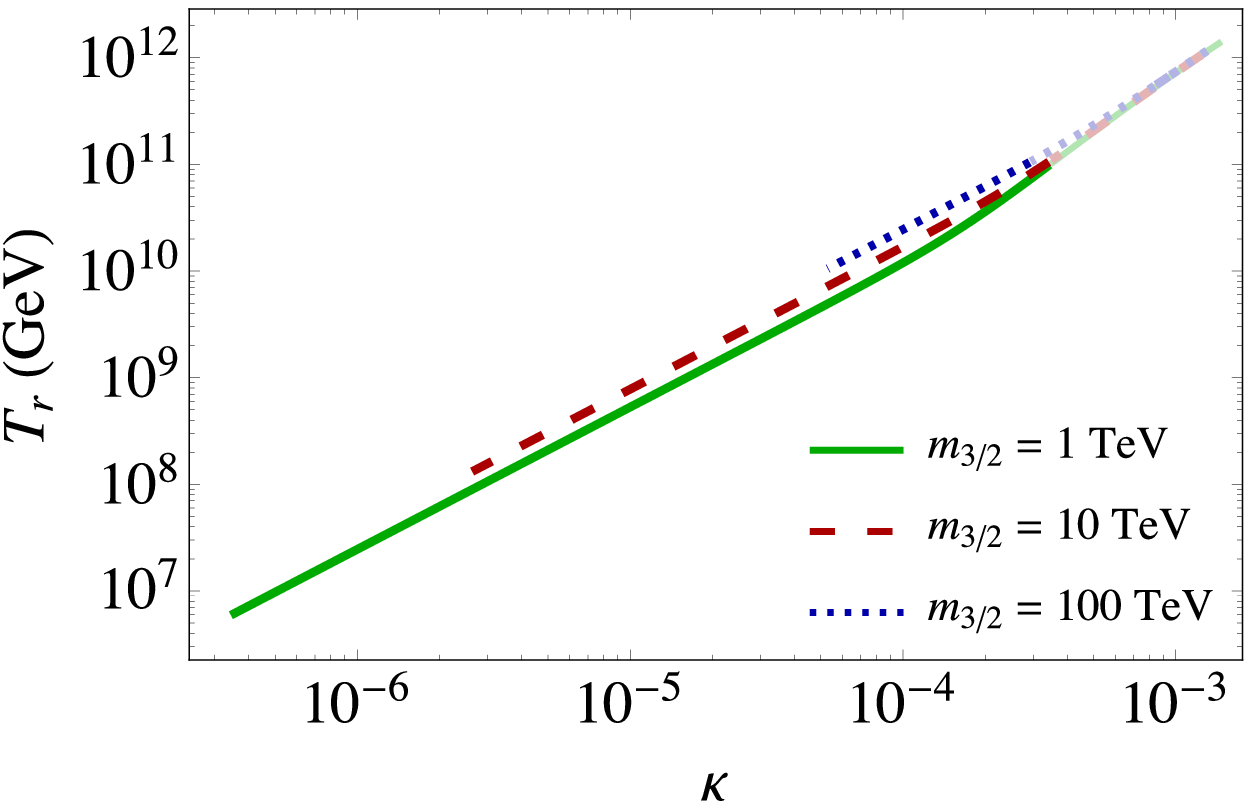}
\includegraphics[scale=0.6725]{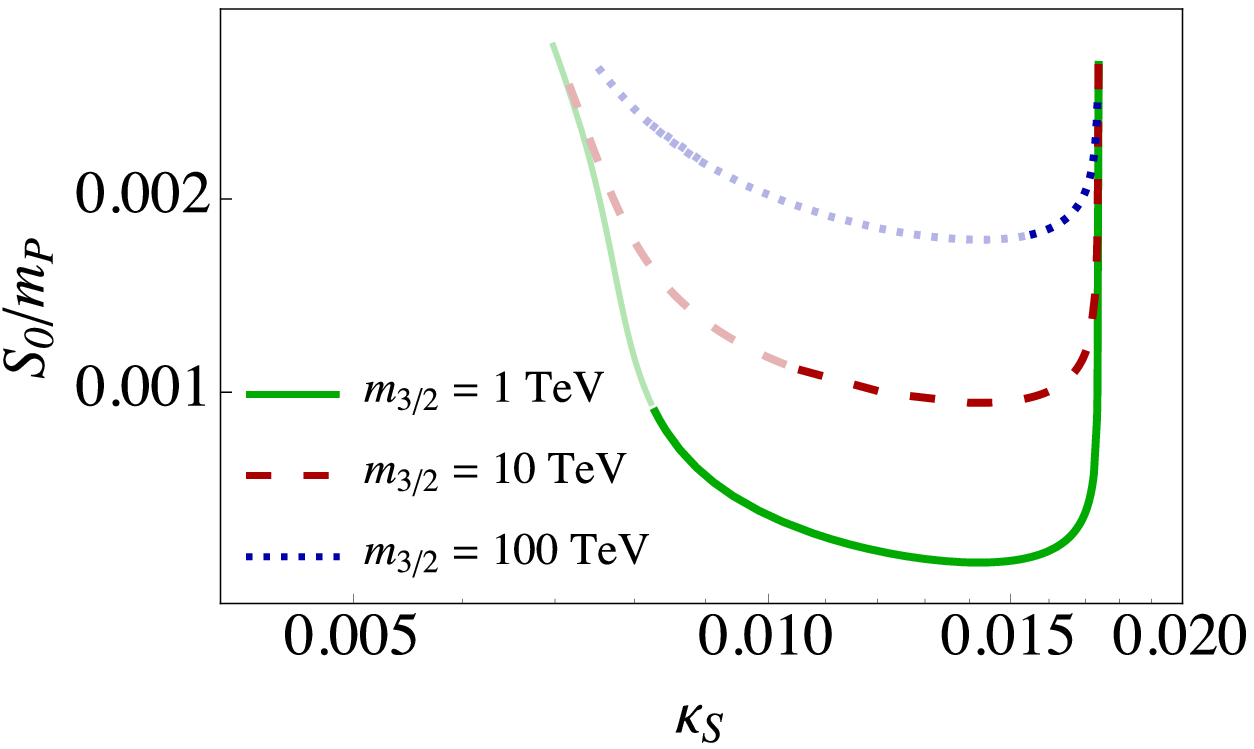}
\includegraphics[scale=0.61]{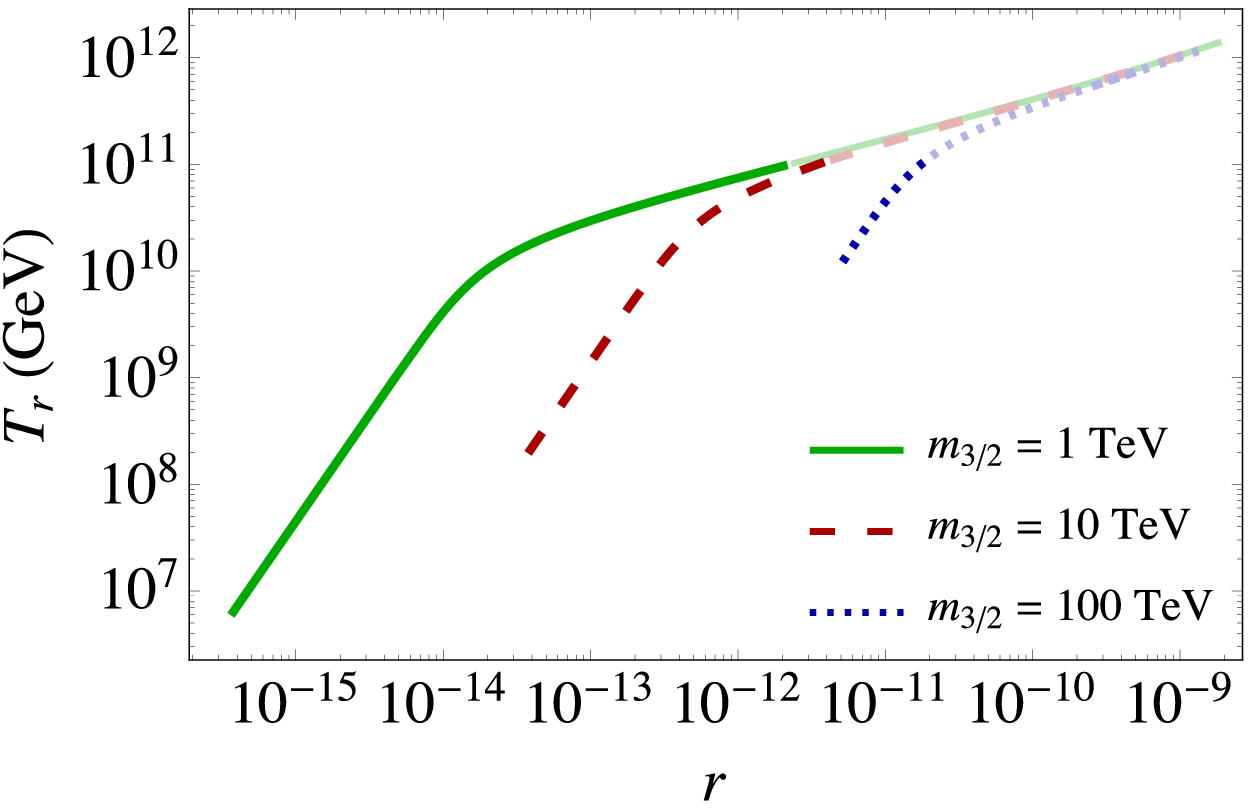}
\caption{\label{fig:CSBMkall} The symmetry breaking scale $M$ and $T_r$ versus $\kappa$ (upper panel) and $S_0/m_P$ versus $\kappa _S$ and $T_r$ versus $r$ (lower panel) are plotted for $m_{3/2}=1$ (solid-green), 10 (dashed-red), and 100 (dotted-blue) TeV with $\gamma=2$, $n_s=0.9655$, and holds for $-1 \lesssim \kappa_{SS}\lesssim 1$. The cosmic string bounds $M_{string}$ are shown by the grey dotted-dashed curve. The faded regions of the curves represent the reheat temperature $T_r\gtrsim 10^{11} \text{ GeV}$.}
\end{figure}

Cosmic strings arise from the breaking of  $U(1)_{B-L}$ at the end of inflation. The observational bounds on these strings are given in terms of the dimensionless quantity $G_N\mu_s$, which characterizes the strength of the gravitational interaction of the strings. Here $G_N = 1/8 \pi m_P^2$ is Newton's constant, and $\mu_s$ denotes the mass per unit length of the string. A recent Planck bound  on $G_N\mu_s$, using field theory simulations of the Abelian-Higgs action, is given by \cite{Ade:2013xla, Ade:2015xua}
 \begin{equation}
  G_N\mu_s \lesssim 2.4 \times 10^{-7}.
 \end{equation}
 For local cosmic strings, $\mu_s$ is given by
 \begin{equation}
 \mu_s= 2\pi M^2 \epsilon(\beta),
 \end{equation}
 where,
 \begin{equation}
 \epsilon(\beta)=\frac{2.4}{\log[2/\beta]} \text{  for  } \beta=\frac{\kappa^2}{2g^2}<10^{-2}. 
 \end{equation}
 For $g=0.7$ and using the above equations, we calculate the Planck bound on the symmetry breaking scale $M$ and denote it by $M_{string}$.
  
 In  Fig.~\ref{fig:CSBMkall}, we display the cosmic string bounds on the prediction of $U(1)_{B-L}$  $\mu$-hybrid inflation for $m_{3/2}=1, 10$, and $100$~TeV.  The cosmic string bounds directly constrain  $M$ and $\kappa$ as shown in the upper-left panel of Fig.~\ref{fig:CSBMkall}. This, in turn, puts bounds on the reheat temperature, especially for larger values of the gravitino mass.  The cosmic string bounds also restrict the value of $S_0$ (with $0.008 \lesssim \kappa_{S}\lesssim 0.0173$ ), which, in turn, constrains the large $r$ values, as shown in the lower panels of Fig.~\ref{fig:CSBMkall}. For further clarity the values obtained for $\kappa$, $r$, $M$, and $T_r$ constrained by the cosmic string bounds are given in Table~\ref{tab:table3}. For  $1$~TeV gravitino mass a wide range of reheat temperature $10^{6}-10^{11}$~GeV will yield observationally acceptable values of the cosmic string tension $\mu_s$. With the increase in gravitino mass, the allowed phase space becomes more constrained. It is important to note that the allowed range of $r$ permissible by cosmic string bounds is highly suppressed and unlikely to be observed in future experiments like PRISM and LiteBIRD. However, if we avoid the cosmic string bound in some way (for instance, by employing the shifted hybrid inflation \cite{Jeannerot:2000sv} or by using $F_D$-term hybrid inflation \cite{Garbrecht:2006az}), then the range of $r\lesssim10^{-6}-10^{-3}$ mentioned earlier is testable in the foreseeable future.
\begin{table}
\caption{\label{tab:table3}The values of $\kappa$, $r$, $M$ and $T_r$ allowed by the Planck 2015 cosmic string bound $G_N \mu_s \lesssim2.4\times 10^{-7}$ for $m_{3/2}=1$, 10, and 100~TeV for fixed values of $n_s=0.9655$ and $\gamma=2$.}

\begin{ruledtabular}
\begin{tabular}{ccccc}
$m_{3/2}$ (TeV)&$\kappa$&$r$&$M$ (GeV)&$T_r$ (GeV) \\
\hline
    1&$3\times10^{-7}-3\times10^{-4}$&$3\times10^{-16}-2\times10^{-12}$&$8\times10^{15}-2\times10^{15}$ &$6\times10^{6}-1\times10^{11}$\\
  10&$4\times10^{-6}-3\times10^{-4}$&$4\times10^{-14}-4\times10^{-12}$&$7\times10^{15}-3\times10^{15}$ &$2\times10^{8}-1\times10^{11}$\\
100&$5\times10^{-5}-3\times10^{-4}$&$5\times10^{-12}-2\times10^{-11}$ &$6\times10^{15}-4\times10^{15}$ &$1\times10^{10}-1\times10^{11}$\\
\end{tabular}
\end{ruledtabular}
\end{table}

\section{\label{sec:level1}Conclusion}
We have considered a $U(1)_{B-L}$ extension of MSSM in which successful supersymmetric hybrid inflation is realized in conjunction with the resolution of the well-known $\mu$ problem. This is achieved by invoking a nonminimal K\"{a}hler  potential as a well-defined power series beyond the leading canonical terms. The reheat temperature lies below the conventional upper bound of around $10^9$~GeV, and the gravitino and scalar sparticle masses can be of order $1-100$~TeV. Thus split supersymmetry, which appears if the K\"{a}hler  potential is minimal, can be evaded. The upper bound on the reheat temperature obtained from the gravitino problem translates into the upper bound range on the tensor-to-scalar ratio  $r\lesssim10^{-6}-10^{-3}$. This is expected to be observed in future experiments. However, it is important to note that this potentially observable interval alludes to the upper limits of a wider range of $r$ predicted by the model. Generalization of $\mu$-hybrid inflation to symmetry groups larger than $U(1)_{B-L}$ should be straightforward and will be discussed elsewhere \cite{Rehman:2017}.

\section*{Acknowledgments}
This work is partially supported by the DOE under Grant No. DE-SC0013880 (Q.S.).

\end{document}